# A Transient Semantics for Typed Racket

Ben Greenman[a,b], Lukas Lazarek[c], Christos Dimoulas[c], and Matthias Felleisen[b]

a   Brown University, USA

b   Northeastern University, USA

c   Northwestern University, USA

**Abstract**    Mixed-typed languages enable programmers to link typed and untyped components in various ways. Some offer rich type systems to facilitate the smooth migration of untyped code to the typed world; others merely provide a convenient form of type Dynamic together with a conventional structural type system. Orthogonal to this dimension, Natural systems ensure the integrity of types with a sophisticated contract system, while Transient systems insert simple first-order checks at strategic places within typed code. Furthermore, each method of ensuring type integrity comes with its own blame-assignment strategy.

Typed Racket has a rich migratory type system and enforces the types with a Natural semantics. Reticulated Python has a simple structural type system extended with Dynamic and enforces types with a Transient semantics. While Typed Racket satisfies the most stringent gradual-type soundness properties at a significant performance cost, Reticulated Python seems to limit the performance penalty to a tolerable degree and is nevertheless type sound. This comparison raises the question of whether Transient checking is applicable to and beneficial for a rich migratory type system.

This paper reports on the surprising difficulties of adapting the Transient semantics of Reticulated Python to the rich migratory type system of Typed Racket. The resulting implementation, Shallow Typed Racket, is faster than the standard Deep Typed Racket but only when the Transient blame assignment strategy is disabled. For language designers, this report provides valuable hints on how to equip an existing compiler to support a Transient semantics. For theoreticians, the negative experience with Transient blame calls for a thorough investigation of this strategy.

## The Art, Science, and Engineering of Programming

## 1 Two Designs, Two Semantics

Over the past fifteen years, the study of language designs that can mix typed and untyped code has emerged as a focal point of programming languages research. The designs differ in two major directions. The first concerns the expressiveness of types. Migratory typing systems aim to accommodate the programming idioms of the untyped world [43, 46]. Gradual typing systems add type Dynamic to a general-purpose type system, providing an easy way to type any piece of code [35, 36]. TypeScript combines these two ideas in a single language [30].

The second difference is about the run-time enforcement of types against untyped code. Natural systems [29] check types at boundaries with higher-order contract wrappers to guarantee type soundness and complete monitoring [12].[1] Transient systems use first-order checks at boundaries and elimination forms to guarantee a weak form of type soundness [48]. Both systems record blame information to explain failed checks, but employ different ways of collecting information and different reporting strategies [14, 50]. TypeScript does not protect its types at all and consequently cannot assign blame when things go wrong.

Researchers should compare the two main design dimensions within a single platform: (1) the overhead of run-time checks and (2) the usefulness of blame information plus the cost of creating and maintaining it. As for (1), a preliminary investigation for the small, functional core of Typed Racket exposes serious performance differences between the Natural and Transient designs [21], calling for a thorough evaluation. As for (2), the work we report in this paper has enabled a first comparison of the usefulness of blame [28].

This paper reports on the challenge of adapting the Transient semantics of Reticulated Python (section 2) to Typed Racket's (section 3) migratory type system. It addresses both the type integrity checking (section 4) as well as the blame-assignment mechanism (section 5). Implementors of other mixed-typed languages, such as the above-mentioned TypeScript, may use this report to adapt Transient to their own compiler (section 6).

## 2 The Starting Point: Transient and Reticulated Python

Reticulated Python implements a rather direct adaptation of the original Transient formal semantics [48, 50]. Similar to most formal models, Transient comes with a bare bones type system. Reticulated adds a few types, most notably structural types for classes and objects. Reticulated does not, however, add special provisions for common Python idioms, many of which do not fit the mold of a conventional type system.[2]

---

[1] Natural is also known as Guarded [48], Behavioral [9], and Deep [47].

[2] PEP 484 (https://www.python.org/dev/peps/pep-0484 accessed 2021-08-31) describes a set of types that is well-suited to Python; it includes true unions and one-off types such as Reversible.

To reconcile its limited type system with the flexibility of Python, Reticulated relies on type Dynamic with implicit downcasts. In the literature, this combination is advertised as the means for programmers to underspecify the types in a portion of code, add precise types to critical parts, and gradually turn their untyped program into a fully typed one [6, 36]. In practice, type Dynamic also serves as a catch-all that must be deployed when existing code stretches the limits of the type system [23, 50]. Many Python programs do not have a fully-typed Reticulated variant; at best, programmers can search for a *most-static* variant [6] that minimizes occurrences of type Dynamic.

Together with its type system, Reticulated brings two fresh ideas: its strategies of shallow checks and collaborative blame. The key observation for the first is that a weak form of type soundness [21, 50] is within reach as long as the program validates shallow, first-order properties at strategic places. Weak soundness predicts the rough outline of result values rather than full type structure. For example, if an expression has type (Listof (Listof Integer)) and reduces to a value v, then Reticulated guarantees that v is a list and says nothing about the list elements. Reticulated enforces this guarantee with first-order checks at all boundaries to less-precisely-typed code and all value elimination forms.[3] As for the second fresh idea, collaborative blame, run-time checks update a global blame map with information about the values they inspect. Although the blame map conflates all uses of a shared value [22] and requires an unbounded number of entries, it provides valuable help by attributing a failed run-time check to a set of boundaries [28].

The argument in favor of both ideas is to trade guarantees for simplicity and efficiency. In particular, Reticulated does not require contract wrappers—neither to enforce types nor to offer blame information. This paper describes how this trade-off works out for a gradual type system that accommodates untyped idioms.

## 3  The Destination: Typed Racket

In contrast to Reticulated, Typed Racket's migratory type system is designed to accommodate idioms that have emerged through the use of plain Racket [19]; Scheme, its immediate predecessor; and the Lisp family in general. These idioms include set-based reasoning about data, types for variable-arity functions, and the heavy use of first-class values including continuations and classes.

Set-based reasoning gives rise to occurrence typing [44], which turns run-time checks into logical assertions and tracks them through the branches of conditionals and the decisions of higher-order functions such as filter. While the type system for the functional subset of Racket is straightforward—other than occurrence types—the type system for first-class classes [39, 41] separates classes from types to accommodate the widely used mixin and trait idioms [17, 18]. Additionally, Typed Racket has universally

---

[3] Elimination forms extract one value from another; refer to a standard text for details [26].

■ **Table 1**  Implementation overview

|                    | **Transient**                          | **Natural**                                          |
| ------------------ | -------------------------------------- | ---------------------------------------------------- |
| **syntax and types** | Typed Racket (for both)              |                                                      |
| **compiler(s)**    | Shallow Racket and SB Racket           | Deep Racket                                          |
| **checking strategy** | shallow checks at boundaries and at elimination forms | deep contracts at boundaries                  |
| **blames**         | a set of boundaries                    | the first boundary at which a type and a value disagree |
| **guarantees**     | weak type soundness                    | type soundness and complete monitoring               |

quantified types tailored to Racket's functional subset [37] and row-polymorphism for Racket's class-based subset.

These unconventional types present a challenge for the Transient approach. Both first-order checks and collaborative blame must be scaled up, if possible, to ensure soundness and provide useful blame errors.

**Side-by-Side**   Table 1 presents the end result of our implementation effort. The Transient and Natural semantics are present under one roof as different compilers for the Typed Racket syntax and type system. Because of the high costs of Transient blame-assignment, there are three compilers in total:

- Shallow Racket implements the Transient semantics without blame,
- SB Racket (or, Shallow Racket with Blame) implements Transient with blame, and
- Deep Racket is the discerning name for the original Typed Racket compiler.

The table also contrasts key properties of the Shallow and Deep semantics. First, for checking types at run-time, Deep Racket uses tailor-made contracts [38] to check deep behavioral properties including type soundness and complete monitoring, albeit with a high worst-case performance cost [24, 40]. Contracts are exactly the implementation complexity that Transient eliminates. Along the way, however, Transient sacrifices compositional reasoning about types; this loss is reflected in Shallow Racket's guarantees. Similarly, contracts in Deep Racket offer the means for annotating values with precise blame information. SB Racket keeps a blame map on the side instead, but the context-insensitivity of this global map and the permissiveness of shallow checks lead to error outputs that, at least in theory, are imprecise [22].

Figure 1 illustrates the implications of Shallow and Deep Racket on a small program that consists of three modules:

- the top module defines an untyped function that sorts a list of values using a given comparison function to determine the ordering,

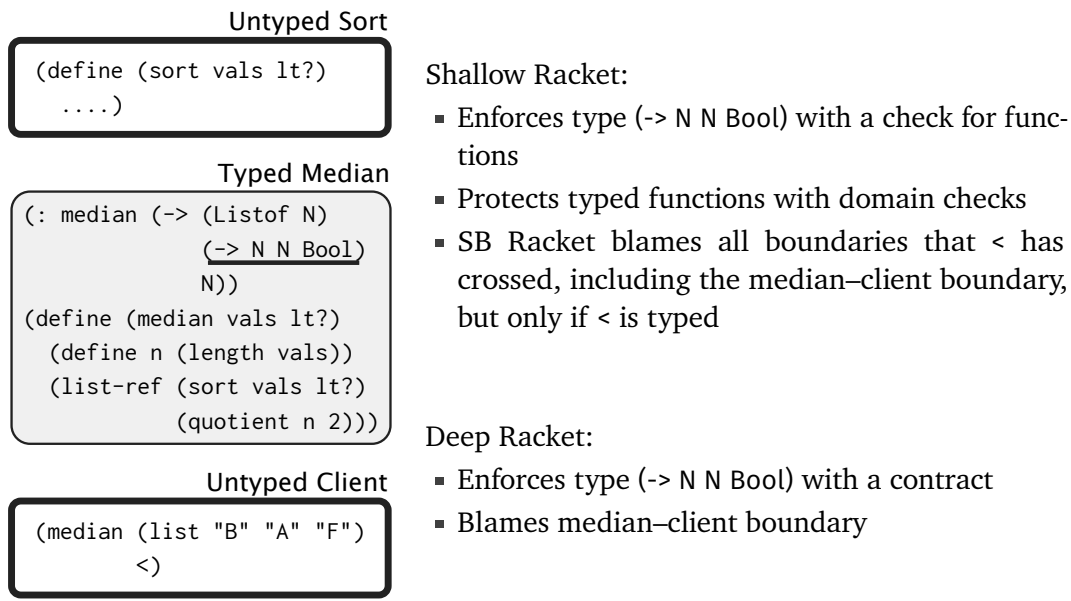

**Untyped Sort**
```
(define (sort vals lt?)
  ....)
```

**Typed Median**
```
(: median (-> (Listof N)
              (-> N N Bool)
              N))
(define (median vals lt?)
  (define n (length vals))
  (list-ref (sort vals lt?)
            (quotient n 2)))
```

**Untyped Client**
```
(median (list "B" "A" "F")
        <)
```

Shallow Racket:
- Enforces type (-> N N Bool) with a check for functions
- Protects typed functions with domain checks
- SB Racket blames all boundaries that < has crossed, including the median–client boundary, but only if < is typed

Deep Racket:
- Enforces type (-> N N Bool) with a contract
- Blames median–client boundary

**Figure 1** Shallow vs. Deep Racket

- the middle module defines a typed median function,
- the bottom *client* module calls median from an untyped context.

The client module also contains a mistake; it invokes median with a list of strings and a comparison function that expects numbers. Both Shallow and Deep Racket can detect this mistake, but under different conditions:

- Shallow and SB Racket can detect an error when the sort module calls the < function, but only if < is declared in typed code. This weak guarantee arises because Shallow enforces the underlined function type with a check that merely looks for a function; thus < flows to the sort function. Only a typed variant of < contains shallow checks that can detect an impedance mismatch.[4] If SB Racket detects such an error, it reports a set containing every boundary that the typed < function has crossed. The boundary between median and client gets included in the set, but, depending on the contents of the global blame map, it may be one among many.

- Deep Racket is guaranteed to detect this error because it enforces the underlined function type with a contract that expects number inputs and halts the program when < receives a string. The contract records blame information that directs the programmer to the faulty median–client boundary.

---

[4] The untyped Racket runtime can detect an error at its own level even when Shallow Racket's checks succeed.

## 4 Bringing Transient Type Integrity Checks to Typed Racket

Adapting the Transient semantics (without blame) to Typed Racket poses two major challenges. The first concerns the kind of checks that the implementation must use to realize the guarantees of transient. The second is about where to insert these checks. The resulting implementation has significantly better worst-case performance than Deep Racket (Natural semantics) across the GTP benchmark suite.[5]

### 4.1 Basic Ideas

Transient uses shallow checks to protect typed code from untyped values. The original model [50] realizes this goal with a type-elaboration pass that puts *tag checks* into three kinds of places:

- Elimination forms get wrapped in a result check. For example, if a cell x has type Ref Int then every read from x gets wrapped in a tag check for integers. The checks ensure safe reads even if untyped code has write access to the cell.
- The entry points of functions check the tags of their arguments. If f is defined in typed code and has the type Int⇒Int, then f gets rewritten to protect its body against untyped callers, which may apply f to non-integer inputs.
- All downcasts from type Dynamic get wrapped in a check.

A tag check is a predicate that is provided by the runtime system and inspects the low-level representation of a value. The garbage-collection tags in SML provide one kind of basis for tag checking [3]. Reticulated implements tag checks with Python's isinstance primitive. Note, however, that Reticulated uses more than tags to enforce some of its types (section 4.2).

**Implementation**    Typed Racket is a macro-based Racket library that injects three major passes into the Racket compiler [42]. Instead of type checking source code directly, Typed Racket first defers to the Racket macro expander and then checks the expanded, kernel-language program. If the program is well-typed, a second Typed Racket pass converts the types at module boundaries into contracts that ensure sound interoperability with untyped code (but do not impede direct typed-to-typed interactions). The final Typed Racket pass is a type-directed optimizer [2] that relies on strong soundness to fine-tune code before it reaches the Racket compiler backend.

Shallow Racket, the adaptation of Transient to Typed Racket, leverages as much as possible of the existing implementation, starting with the expansion pass and the type checker. The second pass is replaced by a *check-insertion* pass that traverses the program to add first-order checks. The final pass reuses all type-directed optimizations that are safe for the weak soundness guarantee. Surprisingly, all but two optimizations meet this safety criterion.

As noted in table 1, this paper refers to Typed Racket's original Natural semantics as Deep Racket and the new Transient semantics as Shallow Racket. To be clear, Shallow

---

and Deep Racket use the same type checker and different methods to moderate typed-untyped interactions. Of the two, Deep Racket is more conservative; it halts some programs that run to completion under Shallow's relaxed semantics.

## 4.2 From Tags to Shapes

Value tags are ill-suited to represent sophisticated types. Indeed, the set of tags and the language of types have radically different and competing design goals. Tags use a small amount of space so that run-time code can cheaply answer frequently-asked questions (is this an integer or a pointer?). Types exist to help programmers design their code and understand it, and need not fit into run-time space constraints. Tag checks suffice for the original Transient semantics only because its type system is limited to integers, reference cells, and functions [50]. Adding a second data structure reveals the growing pains: ML-style tags need not distinguish cells from, say, arrays. Adapting Transient to Typed Racket calls for checks that inspect more than tags.

Shallow Racket thus adapts the notion of *type shapes* from the object-oriented world [7] to represent complex types. A type shape is a first-order property of a type, that is, a property related to the type's top-most constructor or perhaps its immediate constituents. A *shape check* is a predicate that decides whether a value has a specific type shape. It may check a value tag, sign bits, or the mutability settings of a data structure. It may enforce structural properties, for example that an object has at least some $N$ members. A shape may even traverse the full spine of a data structure.

Moving from tags to shapes raises the question of which shapes to enforce. Language designs are beginning to explore this issue. Reticulated uses three kinds of shape: isinstance questions for most types, including lists and strings; unions of tags for numbers, functions, and tuples; and field/method membership checks for classes and objects. Pallene uses tags alone [25]. Grace [34] and SafeTypeScript/Higgs [8, 33] use object shapes alone and thereby benefit from decades of research on optimizing compiler technology [5, 7, 10, 27, 51]. Pyret uses structural checks for fixed-size data and tags for everything else, including lists and functions. Future work is needed to systematically compare these alternatives and to weigh the need for new, or perhaps programmer-controlled, strategies.

### 4.2.1 Representative Shape Checks
Shallow Racket uses shape checks to enforce all first-order aspects of type constructors. This design generalizes the apparent policy in Reticulated. The following examples of types $\tau$ and their shape checks $\lfloor \tau \rfloor$ illustrate how Shallow enforces first-order aspects.

- $\tau$ = (Listof Real) : lists of real numbers
  $\lfloor \tau \rfloor$ = list?
  Accepts any proper list, but not improper lists such as (cons 1 2) and circular chains. The run-time cost depends on the size of input values in the worst case, but the Racket predicate list? caches its results. This shape is the only one in Shallow Racket that may traverse a value of arbitrary size.

■ **Table 2** Deep Racket optimizations and whether Shallow can use them

| Topic | Shallow Ok? | Description |
|---|---|---|
| apply | ✓ | deforest map-reduce expressions |
| box | ✓ | speed up box access. |
| dead-code | ✗ | remove if and case-lambda branches |
| extflonum | ✓ | rewrite math for extended floats |
| fixnum | ✓ | rewrite math for fixnums |
| float-complex | ✓ | unbox and rewrite complex float ops |
| float | ✓ | rewrite math for normal floats |
| list | ✓ | speed up list access and length |
| number | ✓ | rewrite basic numeric operations |
| pair | ✗ | speed up (nested) pair access |
| sequence | ✓ | insert type hints for the runtime |
| string | ✓ | speed up string operations |
| struct | ✓ | speed up struct access |
| vector | ✓ | speed up vector access |

- $\tau$ = (Vector Real Real) : vectors containing two numbers
  $\lfloor \tau \rfloor$ = (λ(v) (and (vector? v) (= 2 (vector-length v))))
  Accepts any vector with exactly two elements. The length constraint lets the type optimizer remove run-time bounds checks when it can determine the size of the offset.

- $\tau$ = (U Real String (Listof Boolean)) : the true union of the three types
  $\lfloor \tau \rfloor$ = (λ(v) (or (real? v) (string? v) (list? v)))
  Accepts either a real number, or a string, or a list; does not check list elements. Wider unions, with $N$ types inside, have shape checks with $N$ predicates.

- $\tau$ = (Class (field [a Natural]) (get-a (-> Natural))) : class w/ two members
  $\lfloor \tau \rfloor$ = (contract-first-order (class/c (field a) get-a))
  Uses the racket/contract library to check properties of a class because there is no public API for these details.

- $\tau$ = (case-> (-> Real Boolean) (-> String String Real)) : overloaded function
  $\lfloor \tau \rfloor$ = (λ(v) (and (arity-includes? v 1) (arity-includes? v 2)))
  Enforces both function arities. In general, case-> accepts $N$ function types.

### 4.2.2 Alternative Choices

Instead of checking all first-order properties of a type constructor, Shallow Racket could have used weaker checks, such as (or/c null? pair?) for lists. Early experiments suggest that simpler checks lead to better performance but noticeably worse error situations; the lax checks make it harder to debug the cause of failures. Though more work is needed to assess usability, we conjecture that full constructor checks are easier to comprehend and to reason about.

### 4.2.3 Optimizations

Shallow Racket's design enables thirteen of Deep Racket's fifteen type-directed optimization topics (table 2). The two remaining optimizations deal with dead function branches and nested pairs. Some of the pair optimizations would, however, be possible if Shallow Racket used deeper checks for nested pair types. Anecdotally, the gradual guarantees provided a helpful framework for reasoning about the correctness of optimizations [36].

### 4.3 Where and How to Inject Shape Checks

Integrating the Transient semantics into an existing compiler requires non-trivial implementation decisions. Many compilers elaborate code written in some surface syntax into an internal kernel syntax [1, 31, 32]. In the specific case of Racket, elaboration is implemented via its macro expander [15] and the type checker is a separate pass behind expansion [11, 45]. The following subsections explain three of the major challenges of integrating Transient into such an elaboration-based compiler.

### 4.3.1 Challenge: Expansion-Introduced Code

An elaboration-oriented compiler operates on the kernel-language representation of the program, but models often formulate properties—such as the insertion of run-time checks—at the source level. The two approaches must be reconciled.

Figure 2 illustrates the distortions of an elaboration with an example. The code at the top-left is a source-level, typed function, which adds up a list of numbers using a for/fold iterator. The code snippet below prescribes how shape checks can protect the function and the loop body. The code on the right shows the kernel-language code into which the check-insertion pass must insert the run-time checks. Instead of a for loop, this kernel code employs explicit recursion; it also contains three elimination forms that the check-insertion pass can ignore (underlined in the figure).[6]

Expansion poses challenges for both performance and correctness. The performance issue is clear. Naively wrapping every expansion-introduced form with a check adds unnecessary costs and can break tail recursion.

The correctness issue is rather surprising. It stems from odd cases in which types underapproximate the behavior of generated code; the bottom line is that enforcing such types causes spurious failures in correct programs. For readers interested in the technical details, figure 3 presents an example.[7] The source code at the top-left asks for every byte in a file and sums these numbers together. A check insertion pass must reason about the expanded version of this source code (bottom-left), in which the for loop has been replaced with a recursive function that interacts with a sequence object. The constructor for the sequence object (make-seq) is overloaded to handle several kinds of input. Unfortunately for the type checker, this constructor can return two kinds of object:

---

[6] The figure is a simplified version of the actual expansion.
[7] This figure is also simplified. The true expansion uses several functions, not an object.

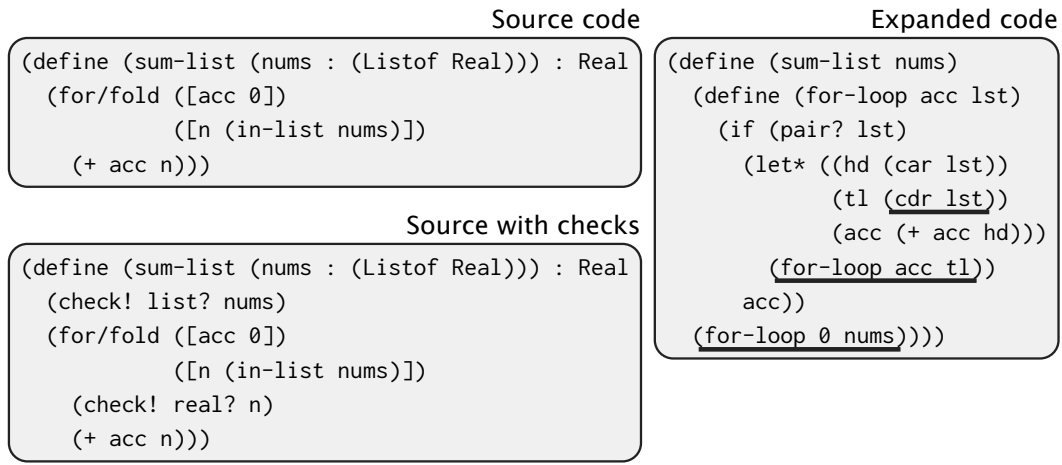

**Figure 2** Typed source code, candidate shape checks, and expanded source code

- For some inputs, the result object encapsulates a sequence of values where every element has the same expected shape.
- For other inputs, the result object encapsulates a sequence with some bad values that an iterator must skip. Files fall into the latter category because every read from a file can return either a byte or an end-of-file token.

The skipping functionality is realized with a conventional Racket idiom. The use-val? field is either #false, pointing to the first case, or a predicate that checks whether the value should be skipped, pointing to the second case. Hence, the expansion in figure 3 works properly on all values, because the conditional inside the loop uses the value retrieved with get-val only if the predicate exists and blesses it.

The existing type system cannot express this conditional reasoning concerning use-val? and get-val.[8] Instead, Typed Racket uses a base type for make-sequence, shown on the right, that under-approximates this reasoning. Specifically, it acts as if the function always returns a homogeneous collection of elements of type T. It is up to the programmer to work around this weakness in the type system with proper uses of use-val?, such as the ones generated by the expansion pass.

At this point, the correctness problem is easy to explain. A naive check-insertion pass would perceive the underlined use of get-val as an elimination form and would wrap it in a shape check. Sadly this check would fail for "skip values" even though the loop body works around such values anyways. In short, the naively inserted check would indicate an error where the expansion code works just fine.

**Implementation**  The compiler passes preceding Shallow Racket's check-insertion pass leave annotations on the kernel-language syntax that differentiate pieces derived from typed source syntax from those injected by the elaboration. Code derived from

---

[8] To precisely type this behavior, the base type for make-seq should use a type-level conditional based on the use-val? member.

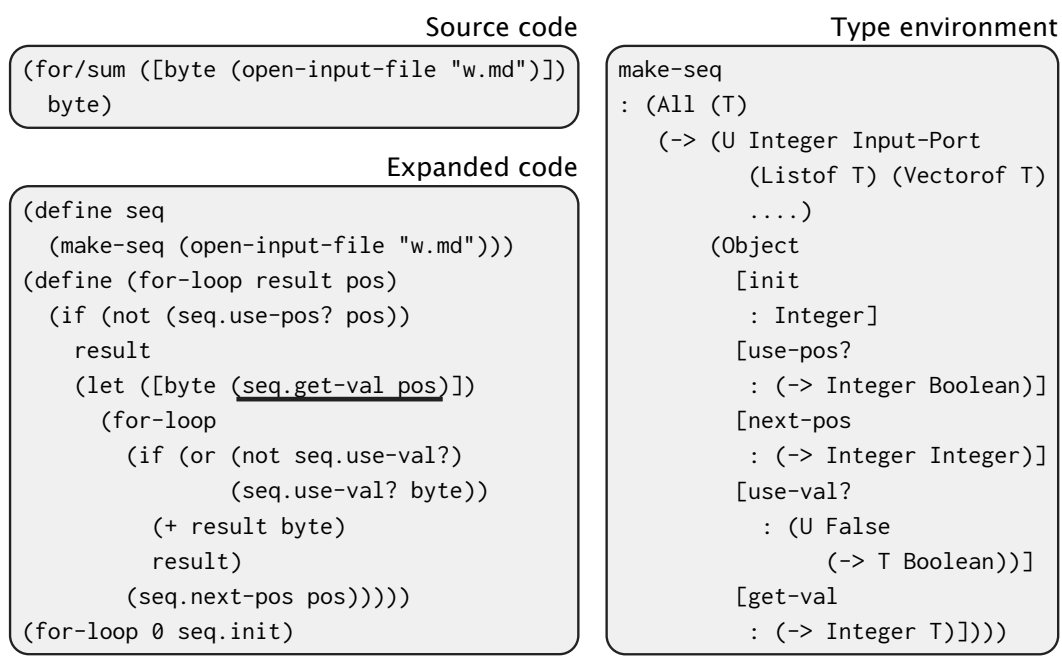

**Figure 3** Underapproximate base type for make-seq

typed syntax is dealt with according to the prescriptions of the Transient model; for
injected code, Shallow Racket does not insert a check. In addition, Shallow Racket
contains special-case patterns to insert correct checks for the sequences used in loops,
for typed subforms in object-oriented code, for exception handlers, and for functions
with optional and keyword arguments. A companion blog post reports the surprising
plumbing challenges of protecting functions.[9]

### 4.3.2 Challenge: Hidden Elimination Forms

Although Racket's type system does not come with type Dynamic, it does have uni-
versal types and occurrence types, both of which create situations similar to implicit
downcasts from Dynamic. A universal type $\forall \alpha. \tau$ is eliminated with a type-level sub-
stitution. An occurrence type introduces type refinements in conditional branches.
For example, the conditional (if (string? x) e0 e1) refines the type of x to String in the
e0 branch.

Many of these type-level deductions require support from run-time checks. Figure 4
presents two examples that illustrate the issues in detail. Both examples are well-typed
but must evaluate to a shape error in Shallow Racket. On the left, typed code imports a
function with an extremely general universal type and instantiates the universal (inst)
to the incorrect String type. A shape check should notice that the identity function is
not a string. On the right, typed code imports a function with an incorrect type that

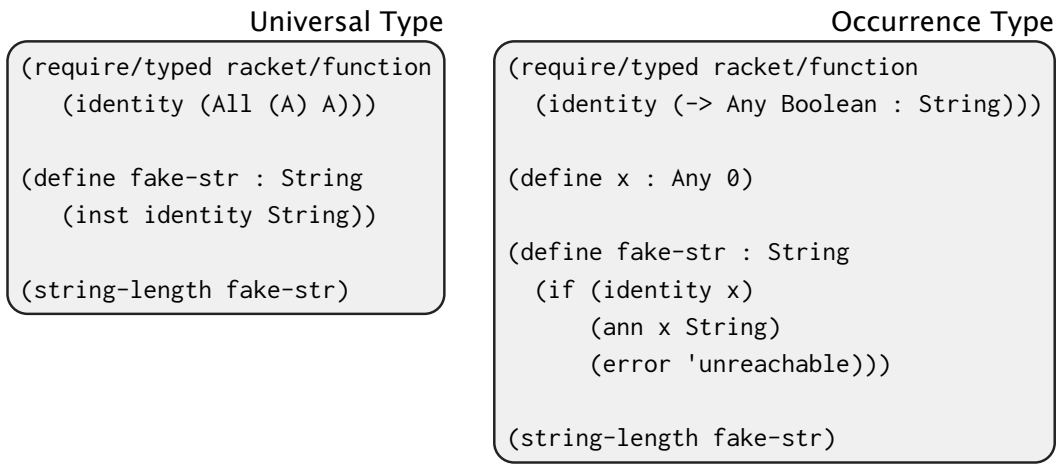

Universal Type

Occurrence Type

```
(require/typed racket/function
    (identity (All (A) A)))

(define fake-str : String
    (inst identity String))

(string-length fake-str)
```

```
(require/typed racket/function
    (identity (-> Any Boolean : String)))

(define x : Any 0)

(define fake-str : String
    (if (identity x)
        (ann x String)
        (error 'unreachable)))

(string-length fake-str)
```

■ **Figure 4**  Two programs that employ type-level reasoning describes the behavior of the string? predicate. A shape check should ensure that x is a string in the first branch of the conditional, indicated with a type annotation (ann).

The challenge is that Racket's elaboration and type checking passes hide source expressions that have no run-time semantics. In the examples, both inst and ann disappear. To properly check such forms, the check-insertion pass would need information from the type checker.

**Implementation**  To avoid changes to the type checker specific to Shallow Racket, the implementation does not communicate type-level eliminations from the type checker to the check-insertion pass. Instead, Shallow Racket takes an overly-conservative approach to some programs. In terms of figure 4, Shallow raises an exception when the identity function tries to enter typed code—at the require/typed form—rather than waiting to see if a misbehavior arises. To be clear, this conservative behavior is no more restrictive than standard Deep Racket, which raises similar exceptions because it lacks contracts to enforce these particular types.

Shallow Racket does support shape checks for universal types whose bound variable appears under a type constructor. For example, the type $\forall \alpha. \alpha \Rightarrow \alpha$ is supported because $\alpha$ appears under the function type constructor. The supported types are a *superset* of those currently supported by Deep Racket's contracts.

### 4.3.3 Challenge: The Cost of Shape Checks

Shape checks can impose a significant run-time cost. Although some of this cost can be reduced in the context of a just-in-time compiler [34, 49], an ahead-of-time compiler such as Racket should take steps to avoid unnecessary checks.

**Implementation**  As mentioned above (section 4.3.1), Shallow Racket is careful to avoid checks around compiler-generated elimination forms. It also comes with a base type environment that says whether a function needs a result check. For example, the map function does not need a check because it always returns a list. The list-ref function, by contrast, requires a check because it may return any kind of value. Though simplistic, this form of reasoning helps reduce costs in an effective manner. More improvements along these lines remain to be investigated. For instance, it may be possible to omit result checks for certain user-defined functions from typed modules.

### 4.4 Transient Enforcement Lowers the Performance Cost

For some mixed-typed programs, the standard Deep Typed Racket runs into serious performance bottlenecks. On the same bottleneck cases, Shallow Racket often exhibits less-severe costs.

This conclusion is based on figure 5; the plots in this figure report the overhead of runtime checks on the GTP benchmark suite version 6.0. Because each benchmark program describes a set of mixed-typed configurations,[10] the curves on each plot count the percentage ($Y$) of all configurations that run no more than $X$ times slower than the untyped configuration. The goal is that 100% of these configurations run with low overhead. The $x$-axis ranges from 1x to 20x overhead relative to the untyped configuration; vertical ticks appear at 1.2x, 1.4x, 1.6x, 1.8x, 4x, 6x, 8x, 10x, 12x, 14x, 16x, and 18x. The $y$-axis counts configurations. The way to interpret a point $(X, Y)$ is that $Y$% of all configurations run at most $X$ times slower than untyped.

The large area under the curves for Shallow Racket suggests better overall performance. Every configuration runs within a 6x overhead in Shallow Racket, and the same configurations often exceed a 20x slowdown in Deep Racket. That said, Shallow is worse than Deep in some configurations of the dungeon and jpeg benchmarks. The reversal happens in configurations that contain a large amount of typed code that infrequently passes values to untyped modules. In these scenarios, the checks spread across typed code slow down Shallow Racket more than the infrequent boundary-crossings hinder Deep Racket.

#### 4.4.1 Experiment Protocol
The data is from a single-user Linux machine with 4 physical i7-4790 3.60GHz cores and 16GB RAM. The machine used Racket v7.8.0.5 (7c90387, before Chez [16]) and a version of Shallow Racket (c074c93) that extends Typed Racket v1.12. Each data point is the result of running one configuration nine times in a row and averaging the speed of the final eight runs.

#### 4.4.2 Cautions
The experimental protocol does not control for VM warmup in a rigorous manner; it merely discards the first run. Warmup may add some noise to the measurements [16]. Such noise is unlikely to affect our conclusion that Shallow Racket is often several seconds, and sometimes several minutes, faster than Deep Racket.

This comparison is biased against Deep Racket because the performance graphs for Shallow Racket omit the cost of blame. Deep Racket automatically collects blame information thanks to the contracts that it compiles types to [38]. A new implementation of these contracts is needed to measure the cost of Deep Racket without blame; refer to prior work for an explanation of the overheads that correct blame entails [13, 20]. That said, we conjecture that a Deep Racket without blame would still fare worse than Shallow because it must create wrappers and traverse large values at boundaries.

---

[10] There are $2^N$ configurations for each program consisting of $N$ modules.

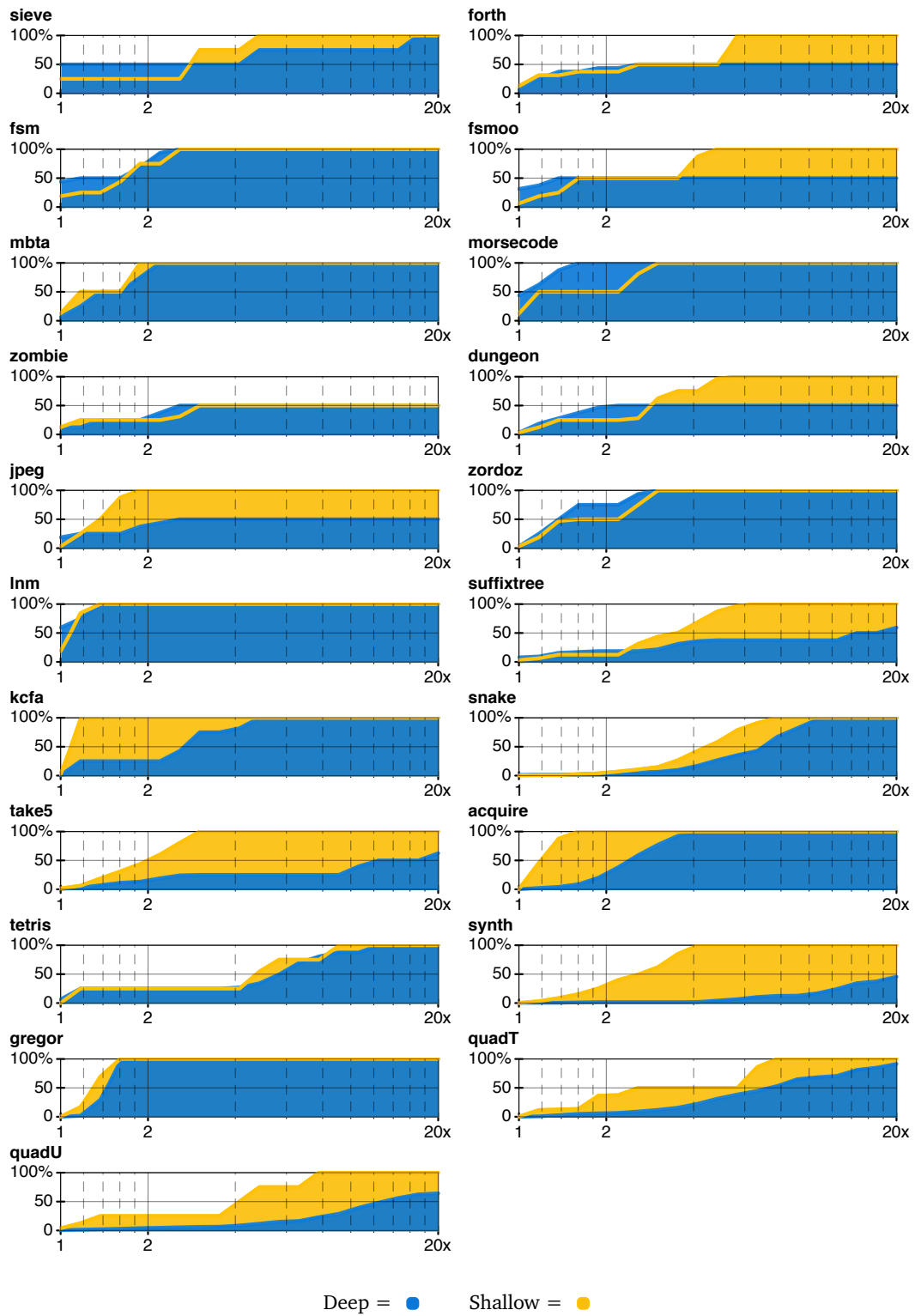

**Figure 5** Performance overhead of Deep and Shallow Racket

## 5 Bringing Transient Blame to Shallow Racket

Adding the Transient blame-assignment strategy to Shallow Racket comes with its own severe challenges. The three most stringent ones concern the integration of existing run-time libraries, the recording of entries in the blame map for complex constructs in the language of types, and the reliance of blame filtering on types.

### 5.1 Adapting the Basics of Transient Blame

The Transient semantics maintains a global *blame map* to implement a blame assignment strategy [50]. This blame map keeps a record of the type casts and checks that occur at run-time. By implication, the map may store an arbitrarily long record for every value in the program.

A blame map is organized as follows. Blame-map keys are heap addresses, which uniquely identify program values. Blame-map values are are sets that may contain two kinds of blame entry; namely, labeled types and pointers:

- A *labeled type* is information distilled from a cast, where the distillation merges a source and target type into one type that is decorated with a source location.
- A *pointer* consists of a blame-map key (heap address) and a context tag. The enclosed key refers to another blame-map value, a *parent* value, and the tag describes an elimination form that was previously applied to the parent.

Consider the example of a typed function f that flows to untyped code. The boundary-crossing is a cast, and the blame map links f to a labeled type. If f has the type (Int -> Int), then the labeled type has the form (Int@A -> Int@-), where A is a source location for the cast and the - states that the result position can never be blamed. Now suppose the same function f gets applied to a value v in untyped code. The blame map gains an association from v to a pointer, which contains both f and the context tag Arg, to remember that v was an input to the function. Subsequent operations may add labeled types and pointers to the sets associated with f and v in the global blame map.

When a Transient check fails, the blame strategy gathers a set of labeled types by computing a transitive closure from the value that failed the check. Each labeled type that is associated with this witness value goes into the set, and each parent pointer starts a recursive search for all labeled types associated with the parent.

To reduce the false positives in these potentially-huge sets of labels, the original Transient model employs a filtering pass [50]. Suppose that a string value v triggers an error and has two parents: a function f that expects a string (matching v) and a function g that expects a number (not matching v). Filtering uses context tags to traverse labeled types, determines the shape of value that each label expects, and checks whether the witness value matches. In this manner, filtering removes the label for f and reports only the problematic g function.

In sum, the success of Transient blame depends on three factors:

1. every typed-untyped boundary in the program *initializes* an entry to the blame map for each new value crossing;

2. every elimination form in typed code must *update* the map with a correct parent pointer and a descriptive action; and

3. the blame strategy must *filter* the witness values via matching against the relevant part of each labeled type.

**Implementation Overview**   SB Racket implements the collaborative blame strategy in a straightforward manner with one significant improvement: it does *not* assume the correctness of types. When a developer retroactively ascribes a type specification to an untyped function, say via require/typed, SB Racket acknowledges that the type may be mistaken [22]. This change in assumptions affects the construction of the blame map in a minor way, but to highlight it, this paper uses a slightly different terminology for blame-map entries. The SB Racket blame map associates heap addresses (keys) to *sequences* with two kinds of entries:

1. A *boundary entry* is essentially a labeled type, but contains source locations for two program components: an untyped client and a type specification. If an error arises, the programmer gets a warning about a bad untyped value under the assumption that the type is correct.

2. A *link entry* is like a parent pointer. It combines a parent pointer with a parameterized action structure (section 5.2.1) to handle a more expressive language of types.

Although SB Racket blames a sequence of boundaries when an error occurs, there is no guarantee that the order has any significance for debugging. The order merely enables reproducible experiments [28].

## 5.2  Engineering Accurate and Precise Blame

The effort of implementing a model for a full-fledged language poses major challenges. Specifically, the above constraints are relatively easy to satisfy for small languages like those used for models, but it is extremely difficult to modify an existing implementation to match them. SB Racket deals with some of the constraints but leaves others for future work.

### 5.2.1  Challenge: Advanced Types Need Advanced Actions

The original Transient model has two elimination forms and three corresponding actions (context tags): Arg and Res actions for function applications, and a Deref action for ref cell reads. The model's elaboration pass has no trouble injecting the correct actions because ref cell reads are syntactically distinct from function applications.

A larger language must support additional elimination forms and must take care not to conflate similar-looking eliminations. If f is a user-defined function, then a

■ **Table 3**   Sample blame actions in SB Racket

| Action Template | Interpretation |
| --- | --- |
| (dom n) | n-th argument to a function |
| (cod n) | n-th result from a function |
| (case-dom (k n)) | n-th argument (of k total) to an overloaded function |
| (object-method (m n)) | n-th argument to method m of an object |
| list-elem | element of a homogeneous list |
| list-rest | tail of a list |
| (list-elem n) | n-th element of a heterogeneous list |
| hash-key | key of a hashtable |
| hash-value | value of a hashtable |
| (struct-field n) | n-th field of a structure |
| (object-field f) | field f of an object type |
| noop | no action; direct link to parent |

call (f x y) requires two distinct argument actions and a link from the result to the f function. If f is an alias for a vector-ref function, however, then the call requires no argument actions and a link from result to the x vector. Clearly, the static type of an identifier (f) should explain which actions are needed.

**Implementation**   At a minimum, an implementation of Transient must look out for special identifiers from the language's run-time libraries, such as vector-ref, to create proper actions in its link entries. SB Racket does so, but nothing more. A precise handling of aliases is left to future work.

The sophisticated Typed Racket type system calls for much more than three atomic actions. In order to render link entries reasonably pragmatic, SB Racket implements sixteen atomic actions and seven forms of *parameterized* actions. Table 3 presents a representative selection. Atomic actions disambiguate operationally-similar data structures. Parameterized actions express a family of elimination forms. For example, the dom action for a function domain is parameterized by a position to handle functions that expect several arguments. The special noop action adds a direct link to track a copied data structure or a user-defined wrapper.

### 5.2.2 Challenge: Trusted Libraries Prevent Initialization

Every language implementation comes with a run-time library, and the compiler (writer) will trust that this untyped library matches its type specification. Deep Racket is no different; it comes with a base type environment that trusts several untyped identifiers. One example is the list-ref function, which comes from an untyped library.

Trusted libraries are problematic for the blame map because they avoid the normal mechanism for initialization. Concretely, a trusted function such as list-ref enters typed code through a back-channel instead of crossing a boundary and submitting to a run-time check. The back-channel ought to initialize the blame map with an entry for list-ref and its type. Performing the initialization is a challenge and may have implications for performance.

**Implementation**  SB Racket does not currently initialize the blame map for trusted functions including list-ref, which means its blame is less expensive and less precise. That said, calls to list-ref do create link entries as specified above (section 5.1). The trouble is that these links may point to list-ref as their parent, especially if list-ref is used in a higher-order manner (section 5.2.1). Such links are dead-ends in the blame map because list-ref is not initialized with a boundary entry.

### 5.2.3  Challenge: Filtering Demands Full Types at Run-Time

The filtering of blame sequences relies on type information. Inspecting types at run-time poses difficulties, though, in an implementation made for separate compilation. Any attempt to serialize the type enviroment must find a way to communicate the identity of generative types from the type checker to the run-time environment. Local type definitions pose another problem; such types exist in a confined scope, which the filtering routine must be able to access within its run-time predicate.

**Implementation**  SB Racket implements filtering, but expects that the process can fail. That is, filtering is understood as a conservative heuristic that, upon failure, does not prune any of boundaries in the blame map.

Notably, the implementation of filtering builds on Typed Racket's protocol for separate compilation. A compiled Typed Racket module contains a serialized copy of its type definitions. The type checker can evaluate such definitions at compile-time to avoid re-checking the module. SB Racket reuses these serialized definitions at run-time to build a type environment in which to do filtering. The reuse is not guaranteed to succeed, but it works often enough to support a heuristic.

### 5.3  Transient Blame Is Extremely Expensive

Table 4 presents the performance of SB Racket and compares it to both Shallow and Deep Racket. The first column lists the benchmarks. The second column reports the worst-case overhead of Shallow Racket across all configurations. The third column reports the worst-case overhead of Deep Racket, also across all configurations. Lastly, the fourth column reports the overhead of SB Racket on the fully-typed configuration— because this configuration contains more blame-map updates than any other. A "TO" entry indicates a time out after 10 minutes. The machine specs and experimental protocol are the same as reported in section 4.4. SB Racket extends Typed Racket v1.12 (49b1005).

The data shows that the collaborative blame strategy adds a huge overhead to SB Racket. Six benchmarks fail to terminate within a generous 10-minute limit. The rest run far slower than the worst case of plain Shallow Racket. Indeed, blame costs more than even the worst case of Deep Racket on seventeen benchmarks.

These bleak results should not come as a surprise. The collaborative blame strategy creates a map of unbounded size and slows down almost every operation with an update. These little slowdowns add up. Deep Racket is slowest only in benchmarks that frequently send large higher-order values across boundaries. Collapsible contracts address this issue for vectors and simple functions but not objects [13], which

■ **Table 4** Blame performance

| Name | Shallow | Deep | SB |
|---|---|---|---|
| kcfa | 1x | 4x | TO (>**540x**) |
| morsecode | 3x | 2x | TO (>**250x**) |
| sieve | 4x | **15x** | TO (>**220x**) |
| snake | 8x | **12x** | TO (>**1000x**) |
| suffixtree | 6x | **31x** | TO (>**190x**) |
| tetris | **10x** | **12x** | TO (>**720x**) |
| acquire | 1x | 4x | **34x** |
| dungeon | 5x | **15000x** | **75x** |
| forth | 6x | **5800x** | **48x** |
| fsm | 2x | 2x | **230x** |
| fsmoo | 4x | **420x** | **100x** |
| gregor | 2x | 2x | **23x** |
| jpeg | 2x | **23x** | **38x** |
| lnm | 1x | 1x | **29x** |
| mbta | 2x | 2x | **37x** |
| quadT | 7x | **25x** | **34x** |
| quadU | 8x | **55x** | **320x** |
| synth | 4x | **47x** | **220x** |
| take5 | 3x | **44x** | **33x** |
| zombie | **31x** | **46x** | **560x** |
| zordoz | 3x | 3x | **220x** |

explains the gap between the fsm and fsmoo benchmarks. The authors conjecture that additional work on collapsible contracts can make Deep Racket run faster than SB Racket on all of the benchmark programs.

### 5.3.1 Cautions

The SB Racket blame map stores full types and source locations. If Shallow instead kept types and source locations in a separate data structure and stored pointers in the blame map, memory overhead would decrease.

The blame map is currently implemented as a weak hash to correctly retain values without inhibiting garbage collection for unreachable entries. This gives slightly better performance than an earlier version that stored only hash codes to reduce the size of each entry. A different implementation technique may further reduce overhead. The fundamental problem, however, seems to be the unbounded space requirements of the Transient collaborative blame strategy.

### 5.3.2 Comparison to Prior Work

These results are far less optimistic than results reported for Reticulated, which state an average slowdown of 6.2x and worst-case slowdown of 17.2x on fully-typed configurations [50]. For comparison, the average slowdown for SB Racket is 133x and the worst-case is 560x, ignoring benchmarks that time out.

These differences are due to two factors: the nature of the benchmarks and the expressive power of the type system. First, the Reticulated evaluation uses rather small programs from the pyperformance suite. Three benchmarks focus on numeric computations; since the blame map does not track primitive values, adding blame adds almost no overhead. Four others have since been retired from the Python suite because they are too small, unrealistic, and unstable.[11] The remaining benchmarks suggest that costs can grow without bound in larger programs. Indeed, a Reticulated version of the sieve benchmark runs in 40 seconds normally and times out after 10 minutes with blame.

Second, as mentioned in section 2, Reticulated uses type `Dynamic` as a catch-all to work around its lack of precise types for Python idioms. This catch-all can even appear as the result of type inference. For example, if an object field begins with the default value `None` and is later initialized to a string, the field must have type `Dynamic`. Polymorphic functions such as range also require type `Dynamic`. These imprecisions significantly reduce costs because interactions among dynamically-typed values do not update the blame map.

## 6   Onward

The experience of constructing Shallow Racket suggests performance benefits yet casts doubt on its potential as a debugging tool for impedance mismatches between typed and untyped components. It thus points to three major pieces of future work.

First, the performance of Shallow Racket is still poor in an absolute sense. One direction for improvement is to develop a static analysis that identifies redundant run-time checks. The Reticulated team [49] reports promising results for one such analysis under a closed-world assumption. Another direction is to equip Shallow Racket with the Pycket just-in-time compiler [4]. The Grace and Reticulated teams have also attributed improvements to tracing JIT compilation [34, 49]; however, the former's semantics does not seem to guarantee soundness and the latter may benefit from the inexpressive Reticulated type system.

Second, the performance of Shallow Blame is clearly unacceptable, worse than the contracts of Deep Racket in almost all cases. The problem seems due to heavy memory usage; it calls for a thorough investigation of the trade-off between the precision of blame and performance. An entirely different blame strategy may be needed. Concurrent to this report, the authors have developed a framework that can measure the usefulness of new strategies by simulating a rationally-acting programmer who uses blame to debug a mutated program [28].

Third, the authors experience with programming in Shallow Racket reveals a usability challenge; namely, programs that appear incorrect can run without error because shape checks arise from local uses rather than type definitions. Consider a typed module that declares three items: a class Animal, a subclass Bear, and a function to-bear with the type (Integer -> (Instanceof Bear)). Suprisingly, there is no guarantee that calls to to-bear return instances of the Bear class. If the function is defined in untyped code, it all depends on where such calls appear. In untyped code, perhaps via (map to-bear ns), the function may return any result. In typed code, an upcast may weaken the result type to an instance of Animal and consequently weaken the run-time check—because Shallow Racket does not wrap the function to enforce the Bear annotation. Language designers must find a way to communicate these and other surprising Transient behaviors [47] to programmers.

---

[11] https://pyperformance.readthedocs.io/changelog.html accessed 2021-08-31

In sum, this paper should help other designers of gradually typed languages that wish to adopt the Transient semantics to add some type integrity to a gradually typed language. A good example would be TypeScript [30] whose type system is rather close to Typed Racket's. In contrast to Racket, a TypeScript implementation with its underlying just-in-time compiler may not suffer much from shallow run-time checks.

**Acknowledgements**   Felleisen and Greenman were partly supported by NSF grant SHF 1763922. Greenman also received support from NSF grant 2030859 to the CRA for the CIFellows project. Thanks to Robby Findler and Northwestern PLT for their valuable feedback at various stages of this work. This paper was supported in part by the NSF grant CNS-1823244.

## About the authors

**Ben Greenman** (benjaminlgreenman@gmail.com) began developing Shallow Typed Racket as part of his PhD research at Northeastern University. He is currently a postdoc at Brown University.

**Lukas Lazarek** (lukas.lazarek@eecs.northwestern.edu) is a PhD student at Northwestern University.

**Christos Dimoulas** (chrdimo@northwestern.edu) is an assistant professor at Northwestern University. Before that he worked as a research scientist at Harvard for several years. His research focuses on the theory, practice, and pragmatics of software contracts.

**Matthias Felleisen** (matthias@ccs.neu.edu) is Trustee Professor at Northeastern University. He has conducted research in programming languages for 40 years, on topics ranging from fundamental theory to practical systems and their evaluation.